\documentclass[ISTS  ]{tjsass} 

\usepackage{graphicx, color}
\usepackage{hyperref}
\RequirePackage{color}
\RequirePackage{multicol}
\RequirePackage{amsmath}
\RequirePackage[varg]{txfonts}
\RequirePackage{bm}
\RequirePackage{array}
\RequirePackage{tjsasscite}
\usepackage[hang]{footmisc}
\newcommand{\bhline}[1]{\noalign{\hrule height #1}}

\pubyear{2017}
\bookvolume{14}
\bookissue{ists31}
\session{d} 
\titleheadertrue
\receiveddate{June 21st, 2017}%

\title{Science Objectives of the Ganymede Laser Altimeter (GALA) for the JUICE Mission}
\subtitle{}
\author{\NAME{Jun}{KIMURA},\thanksNum{1)}\CorresAuthor{junkim@ess.sci.osaka-u.ac.jp} 
\NAME{Hauke}{HUSSMANN},\thanksNum{2)}
\NAME{Shunichi}{KAMATA},\thanksNum{3)}
\NAME{Koji}{MATSUMOTO},\thanksNum{4)}
\NAME{J\"{u}rgen}{OBERST},\thanksNum{2)}
\NAME{Gregor}{STEINBR\"{U}GGE},\thanksNum{2)} 
\NAME{Alexander}{STARK},\thanksNum{2)} 
\NAME{Klaus}{GWINNER},\thanksNum{2)} 
\NAME{Shoko}{OSHIGAMI},\thanksNum{5)}
\NAME{Noriyuki}{NAMIKI},\thanksNum{6)}
\NAME{Kay}{LINGENAUBER},\thanksNum{2)}
\NAME{Keigo}{ENYA},\thanksNum{8)}
\\
\NAME{Kiyoshi}{KURAMOTO},\thanksNum{7)} and
\NAME{Sho}{SASAKI}\thanksNum{1)}}
\thanksOrg{Department of Earth and Space Science, Osaka University, Osaka, Japan}
\thanksOrg{Institute of Planetary Research, DLR, Berlin, Germany}
\thanksOrg{Creative Research Institution, Hokkaido University, Hokkaido, Japan}
\thanksOrg{RISE Project Office, National Astronomical Observatory of Japan, Iwate, Japan}
\thanksOrg{Center for Computational Astrophysics, National Astronomical Observatory  
of Japan, Tokyo, Japan}
\thanksOrg{RISE Project Office, National Astronomical Observatory of Japan, Tokyo, Japan}
\thanksOrg{Department of Cosmosciences, Hokkaido University, Hokkaido, Japan}
\thanksOrg{Institute of Space and Astronautical Science, Japan Aerospace Exploration Agency, Kanagawa, Japan}
\begin{abstract}

Laser altimetry is a powerful tool for addressing the major objectives of planetary physics and geodesy. 
Through measurements of distances between a spacecraft and the surface of the planetary bodies, it can be used to determine the global shape and radius: global, regional, and local topography: tidal deformation: and rotational states including physical librations.
Laser altimeters have been applied in planetary explorations of the Moon, Mars, Mercury, and the asteroids Eros, and Itokawa.
The JUpiter Icy Moons Explorer (JUICE), led by European Space Agency (ESA), has started development to explore the emergence of habitable worlds around gas giants.
The Ganymede Laser Altimeter (GALA) will be the first laser altimeter for icy bodies, and will measure the shape and topography of the large icy moons of Jupiter, (globally for Ganymede, and using flyby ground-tracks for Europa and Callisto).
Such information is crucial for understanding the formation of surface features and can tremendously improve our understanding of the icy tectonics. 
In addition, the GALA will infer the presence or absence of a subsurface ocean by measuring the tidal and rotational responses. 
Furthermore, it also improves the accuracy of gravity field measurements reflecting the interior structure, collaborating with the radio science experiment.
In addition to range measurements, the signal strength and the waveform of the laser pulses reflected from the moon's surface contain information about surface reflectance at the laser wavelength and small scale roughness.
Therefore we can infer the degrees of chemical and physical alterations, e.g., erosion, space weathering, compaction and deposition of exogenous materials, through GALA measurements without being affected by illumination conditions.
JUICE spacecraft carries ten science payloads including GALA. 
They work closely together in a synergistic way with GALA being one of the key instruments for understanding the evolution of the icy satellites Ganymede, Europa, and Callisto.
\end{abstract}
\begin{document}
\maketitle

\section{Introduction}
\subsection{JUICE Mission}
The JUpiter ICy Moons Explorer (JUICE) is a spacecraft mission led by the European Space Agency (ESA) that will provide the most comprehensive exploration of the Jovian system, specifically addressing two key questions of the ESA's Cosmic Vision program: 1) What are the conditions for planet formation and the emergence of life? and 2) How does the Solar System work? \cite{grasset13}

The overarching theme for JUICE is the emergence of habitable worlds around gas giants.
The icy Galilean moons of Jupiter -- Europa, Ganymede, and Callisto -- are believed to contain global subsurface water oceans beneath their icy crusts. 
Ganymede, in particular, the largest moon in the Solar System, can be said to be one of the typical solid bodies along with terrestrial planets in terms of its size and the intrinsic magnetic field originating from the metallic core. 
However, current knowledge provided by previous explorations is extremely limited, since it comes from only several flybys.
JUICE will uncover the whole picture of Ganymede by the first {\it orbiting} in the history around an extra-terrestrial moon. 
The expected new big picture of the origin and evolution of Ganymede will not only be key to unveiling the origin of diversity among the Solar System bodies, but also contribute to our understanding of exoplanets with their broad diversity, possibly including also icy planets.
In addition, the flybys at Europa and Callisto will deepen our understanding of the current state and evolution of the Jovian satellite system.

JUICE is currently planned for launch in May 2022. 
Following an interplanetary cruise of 7.6 years, Jupiter orbit insertion will take place in October 2029.
The spacecraft will perform a 2.5 years Jupiter-orbiting tour including two flybys of Europa at 400 km altitude, multiple flybys of Ganymede and Callisto with a minimum altitude of 200 km.
After these flybys, JUICE will enter into an orbit around Ganymede and stay there for at least ten months.

In this paper, we provide a technical description of the GALA instrument (1.2.), an overview of the Icy Galilean Moons (2.), a summary of the scientific objectives of GALA (3.), and a description of possible synergies with other instruments onboard the JUICE spacecraft (4.). We conclude with an outlook on JUICE mission (5.).

\subsection{The GALA Instrument}
\begin{figure}[!t]%
\centering
\includegraphics[width=80mm,clip]{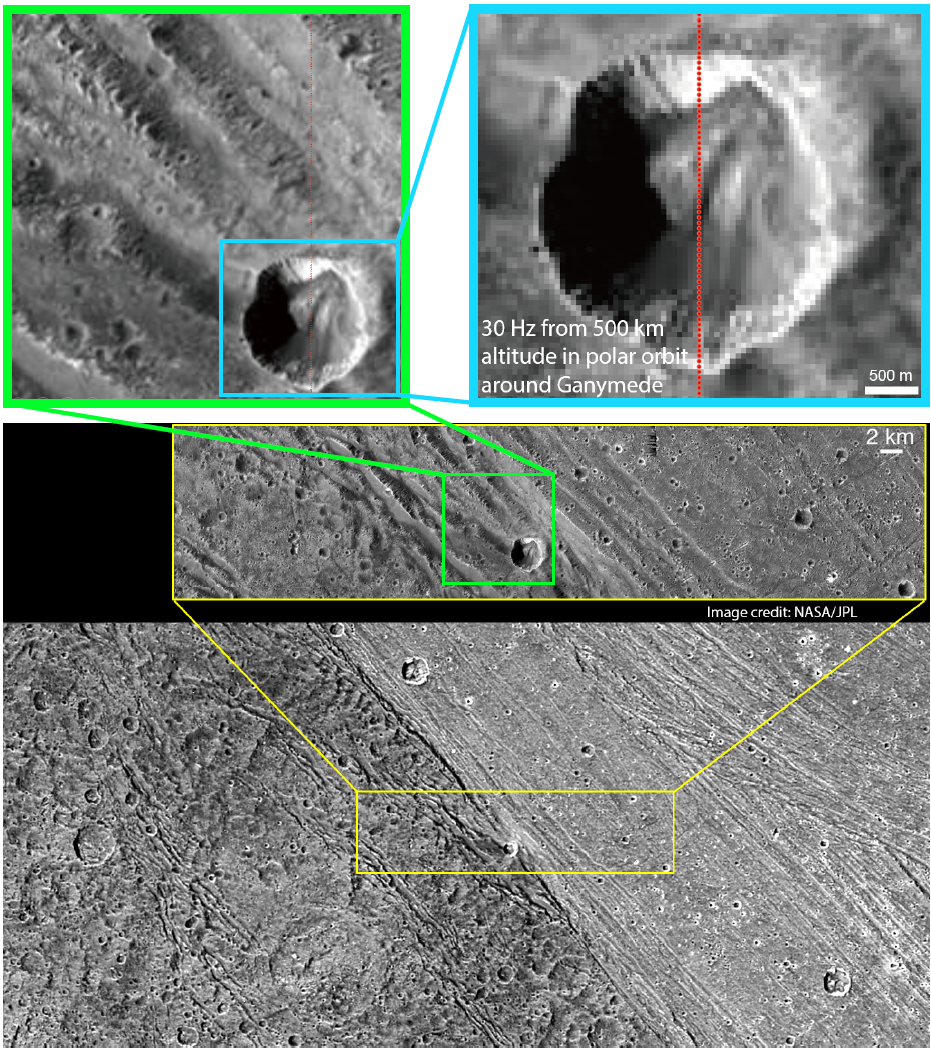}
\caption{Close-up images of Ganymede acquired by the Galileo spacecraft and expected laser footprints of GALA from a 500 km altitude orbit (red dots).}%
  \label{fig:laser_footprint}%
\end{figure}

The Ganymede Laser Altimeter (GALA) is one of ten payloads selected for JUICE mission. 
The GALA uses a straightforward and classical approach of laser altimetry, which means that a laser pulse is emitted at a wavelength of 1064 nm by using an actively Q-switched Nd:Yag laser firing at 30 Hz (in the nominal operation sequence). 
The pulse is reflected from the surface of the target body (surface spot size is 50 m at 500 km altitude orbit (Fig. \ref{fig:laser_footprint})) and is received by a telescope.
The time of flight between the emission of a pulse and the receipt of the reflected pulse is measured, and this flight time is converted to a distance using the light speed.
The GALA will be operated from ranges smaller than 1300 km (depending on the different
albedo values of Europa, Ganymede and Callisto) during flybys. 
The primary observation phase at Ganymede is the final circular polar orbit phase with an altitude of 500 km \cite{hussmann13}.

\section{The Icy Galilean Moons Targeted by the JUICE/GALA}
Is there a life elsewhere in the universe?
Pursuing this question is a fundamental part for human beings,
and a clue to the answer of this question may be found on the icy Galilean moons. 
After magnificent achievements of the Galileo and Voyager spacecraft missions, the existence of a thick liquid water layer underneath solid ice crusts has been inferred for Ganymede, Europa, and Callisto. 
Such liquid layers are now understood as subsurface oceans. 
To date, the evidence for oceans is not fully conclusive and mainly based on electro-magnetic induction signals, surface morphology and thermal modeling.
However, each of the icy Galilean moons exhibits unique characteristics in terms of surface geology, interior structure, and evolution that raises our expectations for habitable ocean worlds.
In the following, we summarize our current knowledge for the three icy Galilean moons, Europa, Ganymede and Callisto, that will be targeted by JUICE mission.

\begin{table}[!t]
\centering
\caption{Orbital data for the icy Galilean moons \cite{weiss04}.}\label{tbl01}
\begin{tabular}{cccccccl}\bhline{0.8pt}
Satellite  & Orbital  &  Orbital & Orbital   &  Eccent- \\
   & Semimajor & Semimajor     & Period   &  ricity  \\
   &    Axis  &    Axis      &          &         \\
   &    (km)  & (R$_{J}$)   &     (days)  &      \\ \hline
Europa &  671100 &  9.38 & 3.551 & 0.0094 \\
Ganymede &  1070400 & 14.97 &  7.155 & 0.0013 \\
Callisto &  1882700 &  26.33 & 16.69 & 0.0074  \\ \bhline{0.8pt}
\end{tabular}
\end{table}%
\begin{table}[!t]
\centering
\caption{Physical data for the icy Galilean moons.}\label{tbl02}
\begin{tabular}{cccc}\bhline{0.8pt}
Satellite  & Mean  & Mass & Mean \\
 & Radius  &   & Density  \\
 &  (km)   &  (10$^{22}$ kg) &  (kg m$^{-3}$)  \\ \hline
Europa  &  1560.8$\pm$0.5 & 4.8017$\pm$0.000014 & 3014$\pm$5 \\
Ganymede  &  2631.2$\pm$1.7 & 14.824$\pm$0.00003 & 1942$\pm$5 \\
Callisto &  2410.3$\pm$1.5 & 10.763$\pm$0.00003 & 1834$\pm$4 \\ \bhline{0.8pt}
\end{tabular}
\end{table}%

\subsection{Europa}
Europa is covered by a solid water ice shell but is primarily composed of rocky materials, as inferred by its mean density of 3014 kg m$^{-3}$ (Table\ref{tbl01}). 
Gravity field measurements by the Galileo spacecraft can be matched with models of interior that range from an $\sim$80 km thick surficial water shell with a larger Fe+S core to an $\sim$170 km thick water shell with a smaller, pure Fe core (the core size also depends on the density of rocky mantle that exists between the water shell and the core).\cite{anderson98}
Europa's surface is characterized by fractured and locally molten ice,\cite{pappalardo99} and the relative paucity of impact crater suggests a geologically short timescale of resurfacing, typically a few tens of Myr.\cite{zahnle03}
The most striking example is the chaotic terrain, showing irregularly-shaped landforms and surface textures which is typically composed of polygonal blocks of pre-existing surface that has been shifted and loosened by the motion of liquid water beneath, expanded, and then has refrozen into hills and jagged mounds.
These chaotic terrains may be located atop vast lakes of liquid water within the ice shell.\cite{schmidt11}
Underneath the solid ice shell, there must be a global liquid water layer because an induced magnetic field has been detected,\cite{kivelson00} although its depth and thickness are not well constrained. 

\subsection{Ganymede}
\begin{figure}[!t]%
\centering
\includegraphics[width=70mm,clip]{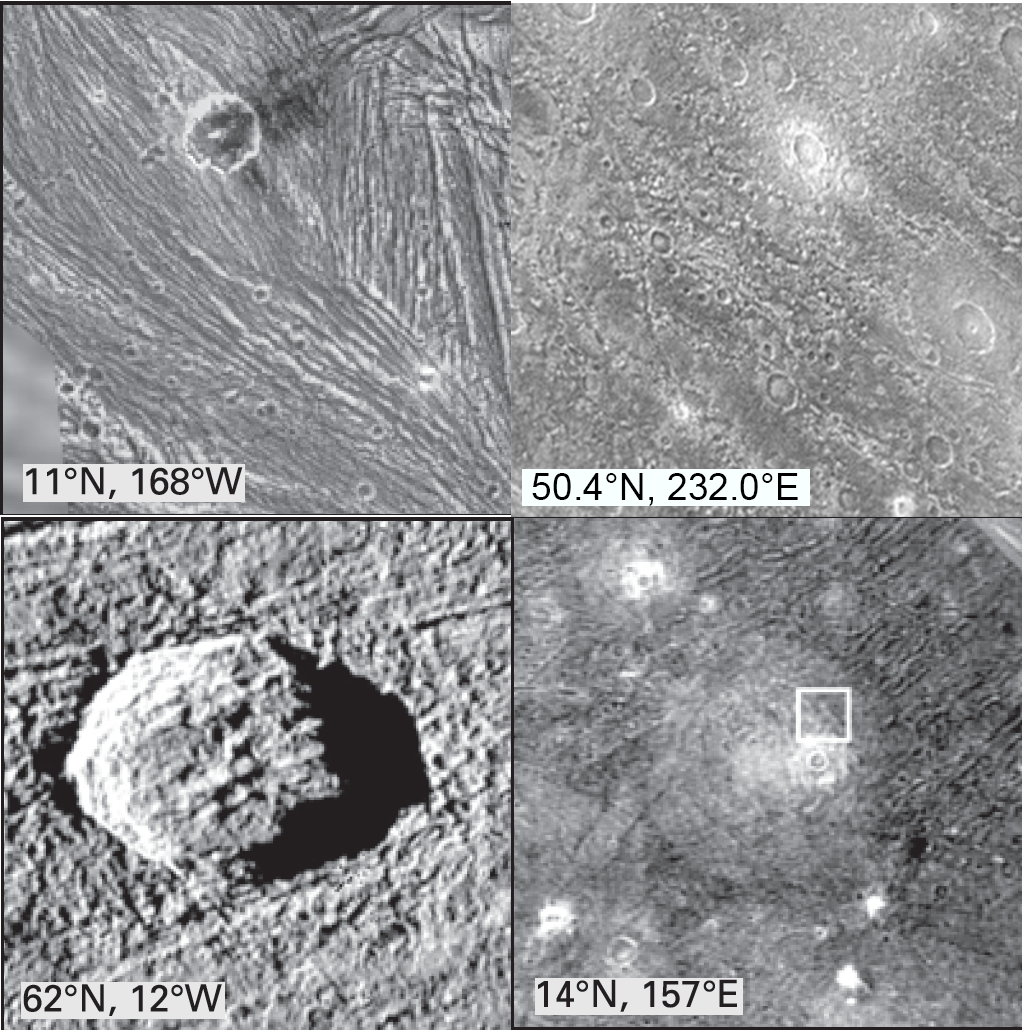}
\caption{Several representative geologic features on Ganymede. Upper-left: Uruk Sulcus (Bright grooved terrain), Upper-right: Lakhmu Fossae (Dark terrain), Lower-left: Achelous impact crater (40 km diameter), Lowe-right: Buto Facula (245 km diameter).}%
  \label{fig:ganymede_features}%
\end{figure}

Ganymede is the largest and most massive satellite not only in the Jovian system but also among all of the planetary satellites in the Solar System, and is the primary target for JUICE mission.
Ganymede is larger than planet Mercury, while its mass is less than half (Table\ref{tbl01}) .
The small value of the moment of inertia factor (0.3105$\pm$0.0028 \cite{anderson96}), derived from gravitational measurements by the Galileo spacecraft, suggests that Ganymede's interior is strongly differentiated.
The strong dipole magnetic field indicates that a (at least partially) liquid iron-rich core exists at the center and that it sustains dynamo activity \cite{kivelson96}.
A possible core radius depends on the relative proportions of major light elements and is not well constrained, but the moment of inertia factor constraints the range of the core radius between 600 km for the pure iron core and 1150 km for the FeS core. 
The thickness of the water layer ranges from 1000 to 1100 km above the silicate mantle.\cite{kimura09,sohl02}
The surface of Ganymede is divided into two principal terrain types, relatively old and dark (heavily impact cratered) terrain, and relatively younger and cross-cutting lanes of bright (typically grooved) terrain (Fig. \ref{fig:ganymede_features}).
Dark terrain comprises about one third of Ganymede's entire surface, and its age is estimated to be older than 4 Gyr based on the crater size and density relationship.\cite{neukum97, neukum98, zahnle03}
The smallest craters are simple bowl-shaped, but it shows transition into more complex morphology with increasing crater size.
Further larger sized craters, morphology changes into central pit, central dome, Penepalimpsests, Palimpsests and Multi-ring structure with its diameter (Fig. \ref{fig:ganymede_features}).
Particularly, vast multi-ringed structures, extending from tens to hundreds of kilometers in length, which are accepted to be fault-bounded troughs formed in response to large impact events into a thin lithosphere in early stage, termed \textit{furrow (fossae)} systems represents dark terrain.\cite{smith79a, smith79b}

The bright terrain covers remaining two thirds of the surface and separate tracts of the dark terrain in the form of wedges.
It is ubiquitously grooved, and indicates episodes of wide-spread extensional tectonism.
Although the bright terrain is fairly heavily cratered, it is clearly younger than the dark terrain as can be seen by truncation relationship with the dark terrain and by the fact that the bright terrain has a lower crater density. 
However an estimated age has large uncertainty because of poor resolution of images and ambiguity of impact flux.
High resolution images revealed that the bright terrain appears to consist of belts of subparallel grooves, and to be the result of tilt-block normal faulting the surface brittle icy layer, while the large-scale topography may be due to necking of the brittle layer over a ductile substrate.
Possible hypotheses for the origin for the bright terrain are global expansion produced by internal differentiation, subsequent heating, and phase transitions in solid ice.\cite{showman97, bland07}
In addition, relatively smooth areas are also present. 
Cryovolcanism has been considered as possible formation mechanism \cite{prockter01} but available data have not provided sufficient resolution and coverage for further enforcing this hypothesis.

Observed inductive response is consistent with a buried conducting shell, probably liquid water ocean, though other interpretations cannot be ruled out.\cite{kivelson02}
The large thickness of water layer makes its structure quite different from that of Europa. 
Due to the high pressure value at the bottom of the water layer, high pressure phase ices must exist below a depth of around 150 km.
Thus, the water layer is subdivided in Ice Ih, Ice III, Ice V and Ice VI with increasing pressure. If a liquid water layer exists, as indicated by the magnetic data, it should be located between the ice I and the high-pressure (HP) ice layer.

\subsection{Callisto}
Callisto is the second largest of the Galilean moons.
The Voyager and Galileo spacecraft revealed that Callisto's surface is fully dominated by impact craters and is one of the darkest among the icy moons with an albedo of about 0.2.
The impact craters show a broad variety of morphologies from small simple bowl-shaped to huge multi-ring structure with increasing size as seen on the dark terrain of Ganymede.
Though an inductive magnetic signature indicating a subsurface liquid water ocean has been detected, its depth and thickness are not constrained and signatures of tectonics and resurfacing are lacking.
A relatively large value of the moment of inertia factor (0.3549$\pm$0.0042 \cite{anderson01}) suggests that Callisto's interior is only partially differentiated, which means the interior is composed of ice and rocks and the amount of rock increases with depth.
The distribution of these components is not well constrained and there is a wider range of interior models between a simple 2-layer model, few hundreds kilometers of water shell over a mixture of ice and rock, and 3-layer model, a small rock+metal core, ice/rock mixed mantle and surficial water shell.
Callisto does not have an intrinsic dipole magnetic field, which is consistent with its lack of a completely differentiated metallic core.
It should be noted that the moment of inertia is obtained assuming a hydrostatic equilibrium state. 
This is mainly because the Galileo flybys were limited to equatorial regions. 
However, non-hydrostatic effects may not be neglected for small or slowly rotating bodies, such as Callisto.\cite{gao12}

\section{Scientific Objectives of the GALA Instrument}
In JUICE mission, the GALA will acquire a first laser altimetry data of an icy body in the Solar System.
During closest approaches of flybys at Europa and Callisto, and in orbit around Ganymede, the GALA will provide range measurements.
These measurements enable us to know the quantitative topography, to monitor the tidal responses, and to measure the albedo and small-scale roughness at surface.
In particular, JUICE spacecraft will be set in a polar orbit around Ganymede in the final phase of the mission.
The 500 km altitude orbit around Ganymede allows for an global coverage of the entire moon.
In previous observations, altimetry data are not obtained and image data has poor resolution; e.g., roughly half of entire Ganymede's surface has been classified as {\it undistinguished} in the global geologic map of Ganymede published by USGS.\cite{map16}
The observations by JUICE will significantly improve such knowledge, then eventually the GALA will provide a characterization of icy tectonics, a constraint on the existence of a subsurface ocean, and understandings of small-scale surface condition and external process of surface alteration.
In the following we will discuss the scientific objectives at Ganymede in detail, the main target of JUICE mission, and briefly summarize the scientific questions addressed during close flybys at Europa and Callisto.

\subsection{Local topography and global shape}
Ganymede and other icy moons are covered with ice on their surface and show many tectonic structures with characteristic morphologies and various spatial scales that are quite different styles from those found in rocky objects.

From range measurements in the 500 km orbit phase, a Ganymede reference ellipsoid will be determined to which regional and local heights measurements can be referred. 
Because of the greater distance from Jupiter, secular tidal effects on the global static shape are expected to be on the order of a few kilometers, much smaller than for Europa or Io. For a typical model Ganymede we would expect a difference of the longest axis (sub/anti-Jovian) to the shortest axis (polar) of about 1.8 km. This deviation from spherical shape is on the same order as elevations of most prominent geologic features on Ganymede, which makes other methods (e.g., limb measurements from imaging data), which can measure only one-dimensional topographic profile, probably less accurate.
The global two-degree shape will be determined by GALA very accurately by fitting a tri-axial ellipsoid to the obtained range measurements.
Deviations from the expected geoid would tell us whether the interior can sustain long-wavelength topography over long timescales. 
Interpretation of such long-wavelength topography would include i) a very rigid outer ice shell that did not yet adjust to the equilibrium state; ii) latitudinal and/or longitudinal variations of ice thickness according to the thermal state of the ice shell and ocean; iii) gravitational influence of the silicate interior on the outer ice shell. Furthermore global shape data will be used to determine center-of-figure/center-of-mass offsets.

Whether lateral variations of topography on the order of thousand km are present on Ganymede is unknown. 
Such topography anomalies superimposed on the degree-2 topography could be caused by density variations in the interior. 
Correlation between gravity and topography would indicate deviations from isostatic compensation with important implications for the thermal state of the lithosphere. 
Therefore, combined gravity and topography measurements are key to understand the thermal evolution of the outer ice shell. Inhomogeneous mass distributions in the ice can be caused by convection, impact remnants buried in the ice, or by large scale topography at the rock/ice boundary. 
Gravity anomalies at scales of a few thousand km on Ganymede have been reported \cite{palguta06, palguta09}. 
Combined gravity and topography data at the given locations could reveal whether the mass anomalies are located deep inside the interior (possibly associated with the rock/ice boundary) or correlated with surface topography and geological terrain. 
Multi-ring impact basins, e.g. the large basin Gilgamesh on Callisto, have similar dimensions. 
The gravity corrected for topography would reveal whether such basins show anomalies associated with mass concentrations.

Topography provides a specific record of the evolution processes and lithospheric structure of a planetary body. 
It can be used to infer lithospheric thickness and, using Yield Strength Envelopes (YSE) \cite{giese08, golombek86}, determining the body's thermal state at the time of (topography) formation. 
Laser altimeter profiles combined with camera data are appropriate to infer the effective elastic thickness of the lithosphere \cite{nimmo02}. 
In order to get such topography information the lateral spacing between the profiles must not be larger than a few km. 
The morphology of the boundary reveals that the trough and the elevated bright terrain flanking ridge were formed by extension leading to the model of lithospheric flexure in this case \cite{nimmo02}.

Thus, local topography will help to understand the formation of geological features that are related to the structure of the outer icy crust, the processes involved in its deformation and the specific fate of water in this layer adjacent to the conjectured internal ocean: on Ganymede, the high strain tectonism probably implied in the formation of grooved terrains \cite{pappalardo98} and possibly cryovolcanism both give indications on the rheology and thermal state of the icy crust.

The coverage of the topography of Ganymede by the GALA instrument is subject to operation time and the spacecraft orbit.
Due to the polar orbit of JUICE the coverage is best at polar regions and worse at equatorial regions. 
The coverage by the GALA after the nominal mission the global shape can be expressed in spherical harmonics coefficients with degree and order 44. 
According to the sampling theorem of Ref. \citeN{driscoll94}, this corresponds to a grid size of 2$^{\circ}$ corresponding to 92 km at the equator.
The global shape of Ganymede will be therefore determined with an accuracy of better than 5 meters.

In the next section, expected scientific objects of surface geology for the targeted moons will be summarized.

\subsection{Geology}
For Ganymede, detailed topographic profiles crossing the linear features of grooved terrains as well as digital elevation models (DEMs), produced from the dense altimetry coverage obtained from the circular orbit phase, will be central for characterizing the tectonic styles and processes associated with groove formation, as well as for mapping regional tectonic systems that form natural boundaries for the analysis of large-scale crustal deformation.
Specifically, the total amount of surface expansion due to quantitative morphology analyses of grooves would be seen by topographic data, and we might see the amount of internal volume change due to temperature change and/or the solidification amount of the subsurface ocean, which will be important constraints for the thermal history of Ganymede.
In addition, identification of small impact craters enable us to estimate the geologic age with higher accuracy.
Finally, information by combining morphology of each geologic unit and their age will bring an essential contribution to understand the style of the icy tectonics and its evolution.

Ideally, this will benefit also from context information from high-resolution images of the camera experiment, JANUS, as well as from stereo-image derived DEMs.
The topographic database derived from stereo imaging from the Galileo and Voyager datasets is very limited. 
The available data suggest that Ganymede's surface is subdued as observed also for Europa.

Laser altimetry for the dark terrain and the impact crater enable us to find quantitatively the depth and diameter of impact craters, which reflect the structure of icy lithosphere and the interior thermal state.
Crater morphology changes with time which means that the depth of the crater floor decreases, called as \textit{relaxation}, thus topography data by the GALA will be a key to constrain the moon's thermal history.
Relaxation of the impact crater changes stress field and induces faulting within and around the crater, and the morphology of faults (radial or concentric) depends on the lithosphere thickness \cite{kamata14}.
Therefore, measurement of depth--diameter ratio and analyses of fault structure for a crater large enough for significant relaxation will be important for understanding Ganymede's thermal history.
Accurate measurement of the depth--diameter ratio for small craters, which is considered to be hardly relaxed, is also necessary to estimate the relaxation degree of large craters, e.g., Tashmetum and Herchef crater.

\begin{figure}[!t]%
\centering
\includegraphics[width=70mm,clip]{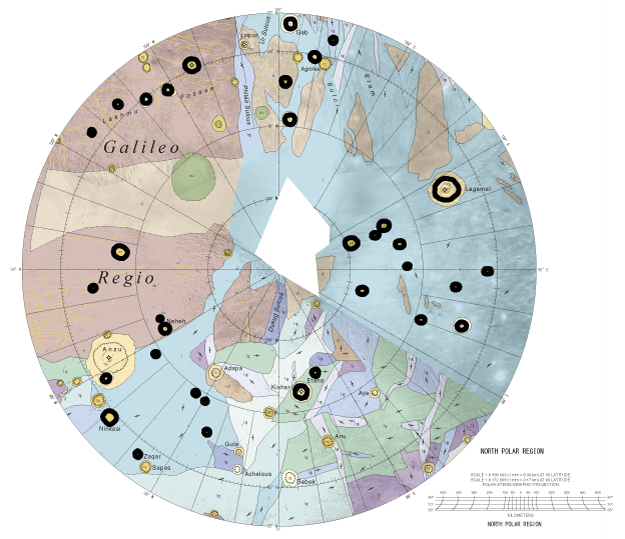}
\caption{Geologic map of Ganymede's north polar region overlapped with surface image taken by Galileo spacecraft. Black circles indicate the impact caters on a single geologic unit. Modified from Collins et al., 2013 \cite{map16}}.%
  \label{fig:craters_ganymede}%
\end{figure}

Because JUICE spacecraft will enter the pole--to--pole orbit around Ganymede in its final mission phase, the GALA's altimetry footprints will have a higher density in a high latitude area.
Thus impact craters located around polar region are preferable to measure small-scale morphology inside a crater with a higher spacial resolution, e.g., Ptah crater (Fig. \ref{fig:craters_ganymede}).
In addition, craters located around the boundary of different geologic units (e.g, between the dark terrain and the bright terrain), or craters deformed by subsequent tectonics (e.g., transected by faults) make estimation of depth--diameter relationship difficult.
Therefore, impact craters located in a single geologic unit not affected by any subsequent tectonics are preferable.

Generally, old dark terrain features seem to have stronger relief than younger, bright terrain features. 
The features with largest relief are domes showing relief up to 2.5 km \cite{prockter10}.
While the requirements from gravity and geophysics are stringent with respect to height resolution but generous in terms of spatial resolution, for geology, high spatial coverage, high spatial resolution and high resolution of topographic heights are all important because of the smaller scale of the tectonic and possibly hydrovolcanic features. 
Combined with global image coverage, the GALA operating at 30 to 50 Hz combined with a laser footprint of 50 m diameter will provide excellent along-track coverage.

For Europa, the GALA will perform altimetry at an altitude less than 1300 km during two flybys.
Europa's diverse geology is unique in the solar system.
Spreading-style crustal deformation is known to be a dominant process of surface formation,\cite{greeley00} while chaos regions and ridges are most certainly associated with melting and melt migration within the icy crust (e.g.\cite{schmidt11}).
The GALA's high frequency mode with laser shots at 50 Hz is appropriate to obtain height profiles with a spot distance down to 20 m when the spacecraft is closer to the surface than about 1600 km. 
Along the two ground tracks several regions are located that are associated with recent activity or possible cryo-volcanism presumably involving liquid water as indicated by smooth plains. In chaotic terrain (e.g., Conamara Chaos) the ice blocks are broken apart, tilted, and shifted within the matrix material which appears to be at different topographic levels.
A fundamental question related to the formation of chaos terrain is whether chaotic regions are elevated regions with respect to the surrounding reference surface. 
The latter would suggest upwelling plumes as the fundamental formation process.\cite{pappalardo99}
An alternative scenario favors liquid water associated with cracks in a thin ice shell.\cite{greenberg00} 
Besides chaotic regions double ridges, bands, smooth plains, pits, spots and domes, and craters would be sites of specific interest along the tracks. 
Again the synergy between laser altimetry, camera, spectrometers and subsurface radar for these targeted observations would provide most valuable data-sets to infer the dynamics and recent (or ongoing) activity in Europa's ice shell.

Whereas Ganymede and Europa show regions that have been modified extensively by geological processes, the surface of Callisto is dominated by craters of various sizes. 
Besides impacts, surface modifications on Callisto are based on weathering processes and erosion. 
Whereas surface roughness measurements are the tool to characterize different degrees of surface degradation, height profiles obtained during flybys would reveal crater morphologies of various types. 
The GALA will address both objectives during the Callisto flybys. 
The latter would be related to impact crater formation and different rheological structure and temperature profiles of the shallow ice at time of formation. 
Specific targets during the flybys at Callisto would include the various types of impact structures from simple craters to largely relaxed structures.
Palimpsests are unique impact structures on Ganymede and Callisto that are devoid of typical crater morphologies such as crater rims. 
Origin hypothesis of such heavily eroded and relaxed impact structures include the following modes of emplacement: (a) extrusion, triggered by impact, (b) fluidized ejecta, (c) dry and solid ejecta. To distinguish between those by profiling the surface and subsurface is a fundamental task of laser altimetry, imaging systems, and subsurface radar.

\subsection{Diurnal tidal responses}
During the mission lifetime, the GALA onboard JUICE spacecraft will conduct multiple measurements at many geographically fixed locations and will provide a time series of surface displacement \cite{steinbrugge15}. 
The temporal changes of the surface displacement are likely to be due to tides raised by the primary, Jupiter, because the orbit of Ganymede has non-zero eccentricity and the interior of Ganymede is not perfectly rigid. Tidal deformation also leads to a temporal change in the gravity field, which is also planned to be measured by the tracking of the spacecraft \cite{grasset13}. 
Since different interior structures lead to different amount of deformation (i.e., a softer interior leads to a larger amplitude of deformation, and vice versa), tidal deformation measurements through laser altimetry (i.e., the GALA) and the radio science experiment (3GM) are key observations to constrain the interior structure of Ganymede. 
In particular, the presence or absence of a subsurface ocean would have a large effect on tidal response. 
Although there are large uncertainties in many parameters describing the interior structure of Ganymede (i.e., the thickness of a subsurface ocean, the viscosity and rigidity of ice), previous theoretical studies of Ganymede's tidal deformation \cite{steinbrugge15, moore03, wahr14} explored only a limited range of parameters. 
We calculate (spherical harmonic) degree-2 tidal response of Ganymede adopting different interior structures. Specifically, we calculate the amplitude and the phase lag of surface displacement and gravity field under a wide variety of parameter conditions.

We use a spherically-symmetric, fully differentiated Ganymede model consisting of an outer water layer, a rocky mantle, and a metallic core. 
The detailed framework of our model is described in Ref.\citeN {kamata16}.
The reference viscosity for the ice shell (i.e., the viscosity at the melting point) and the viscosity of HP ices are free parameters.

The degree-2 tidal response for a spherically symmetric body can be calculated by solving a linear ordinary differential equation system. The non-dimensional values describing the tidal response are called the Love numbers: $h_{2}$ is the normalized surface vertical displacement and $k_{2}$ is the normalized gravity field response. We use a Love number calculation code we developed previously \cite{kamata15}. 
The Love numbers are complex numbers; the absolute value gives the amplitude of deformation while the argument gives the phase shift of deformation due to viscous behavior of the body.

\begin{figure}[!t]%
\centering
\includegraphics[width=70mm,clip]{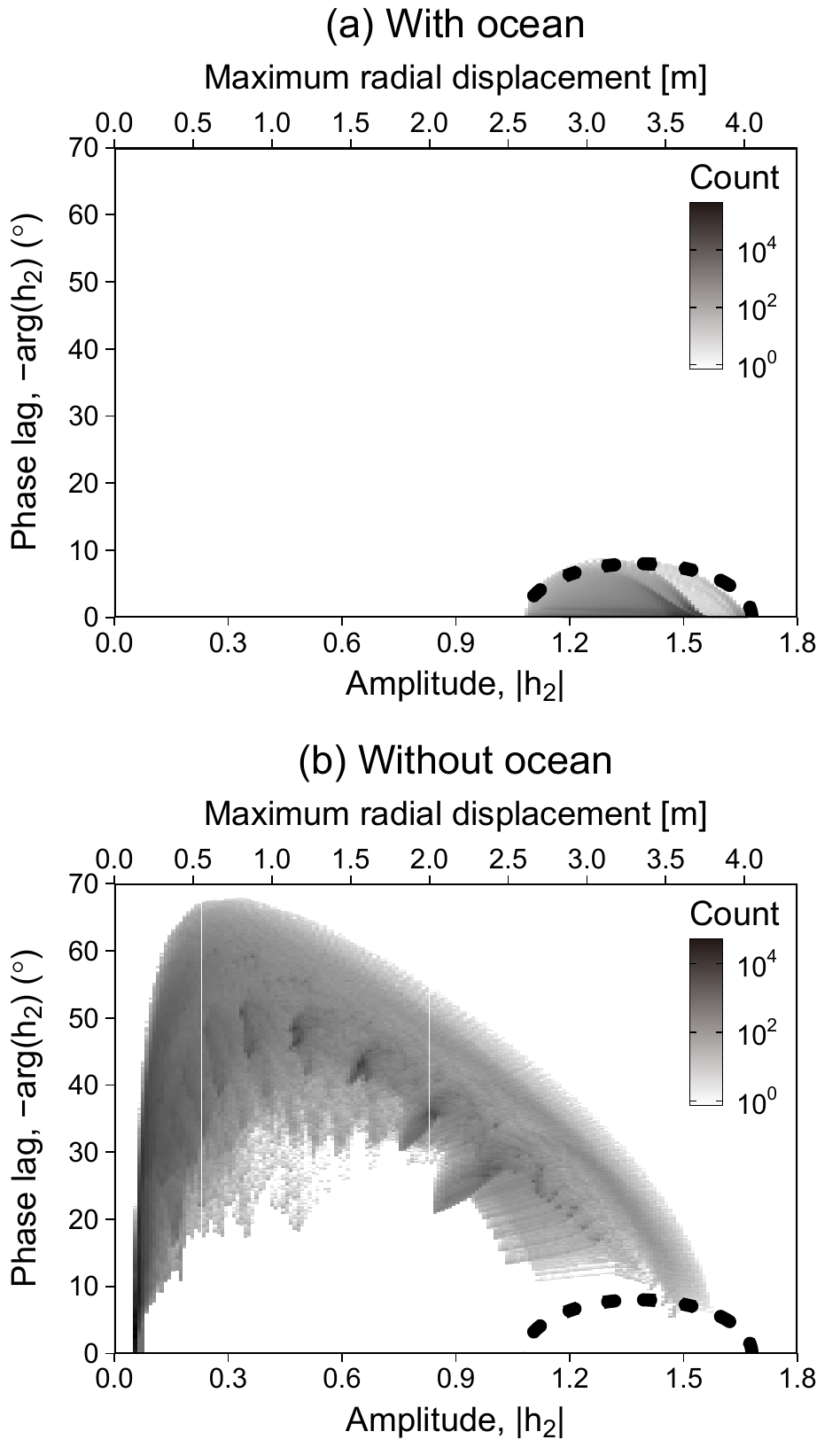}
\caption{Distribution of the Love number $h_{2}$ for (a) cases with a subsurface ocean and for (b) cases without an ocean. The dashed oval represents the distribution of solutions for cases with an ocean.}%
  \label{fig:h2}%
\end{figure}

\begin{figure}[!t]%
\centering
\includegraphics[width=70mm,clip]{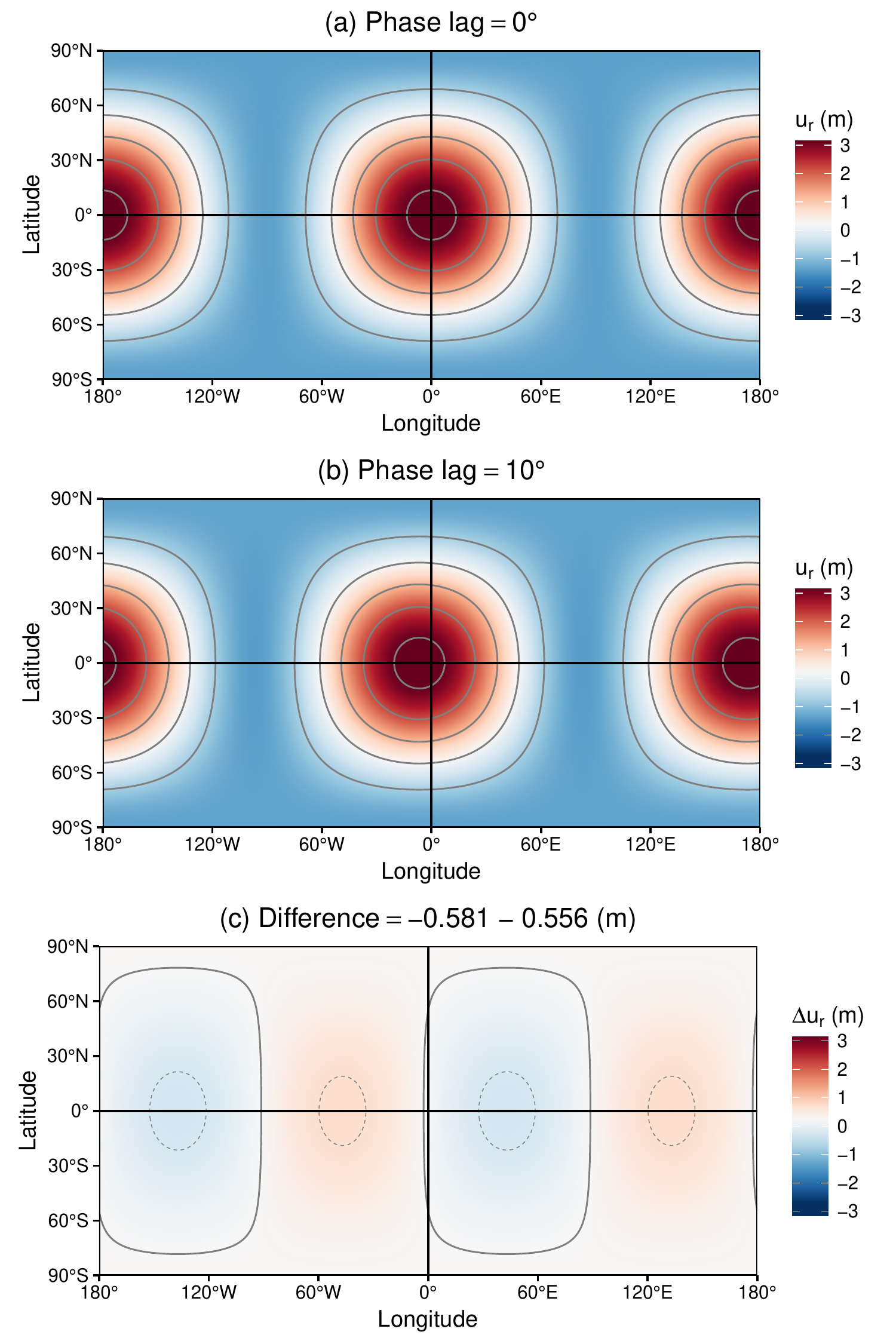}
\caption{
Top: Expected radial displacement on Ganymede at perijove assuming $|h_{2}|=1.45$ and no phase lag. 
Middle: Same as the top panel but assuming a phase lag of 10 degrees. 
Bottom: Differences between above two figures (middle - top). 
Contour intervals are 1 m (solid) and 0.5 m (dashed).}%
  \label{fig:phase_lag_difference}%
\end{figure}

Figure \ref{fig:h2} shows the two-dimensional histogram of $h_{2}$ (radial displacement response) calculated under different parameter conditions (15,537,184 cases).
If there is a subsurface ocean, the absolute value of $h_{2}$ ranges between 1.0--1.7, and the phase lag is \verb|<| 10 degrees.
On the other hand, if there is no subsurface ocean, the former ranges between 0.1--1.6, and the latter can be up to even more than 60 degrees.
Phase lag here is the angle between the position where the tidal potential becomes maximum, i.e. perijove, and the position where the radial bulge becomes maximum on the Ganymede's orbit.
Phase lag of 60 degrees means that the maximum radial budge should occur about 29 hours after passing through the perijove (Ganymede's orbital period is $\sim$ 172 hours).
On the other hand, when the phase lag is 10 degrees, maximum radial bulge would occur about 5 hours after the perijove (Figure \ref{fig:phase_lag_difference}). Because this time difference is much larger than the time resolution of topographic measurements achieved by GALA, we should be able to tell the presence/absence of a large subsurface ocean from the time lag measurements.

This result indicates that one cannot infer the presence of a subsurface ocean in Ganymede solely based on the amplitude of tidal deformation; a large absolute value of $h_{2}$ ($\sim$1.5) does not necessarily requires the presence of the ocean. This is the case when the HP ice layer, in particular the ice III layer, has a low viscosity. If a rheology experimentally determined under a high stress were applicable to the actual HP ice of Ganymede, an extremely low viscosity may be rejected. 
Nevertheless, the rheology of HP ices under a low stress is poorly known. 
Thus, one cannot rule out the possibility that the effective viscosity of the HP ice layer is very low.

We found that this result is quantitatively the same for $k_{2}$ (gravitational response).  
Nevertheless, if we use both the amplitude and phase lag, one can infer the presence/absence of the ocean. If a subsurface ocean exists, a large amplitude ($|h_{2}|$ \verb|>| 1) and a small phase lag (\verb|<| $\sim$10 degrees) is expected. Such a large $|h_{2}|$ can be achieved even if the ocean does not exit, though it leads to a large phase lag (\verb|>| $\sim$10 degrees). Thus, observational constraints on the amplitude and the phase lag of tidal deformation would play a crucial role in inferring the presence or absence of a subsurface ocean in Ganymede.
Again, this is also the case for $k_{2}$.

Simulation studies are necessary to estimate the accuracy of the Love numbers. 
Ref. \citeN{steinbrugge15} presented such a simulation result for $h_{2}$ based on the GALA instrument performance and the current spacecraft operation plan. 
They found that about half a million crossover points are expected to be formed during the nominal mission period of 132 days with realistic operational time of 16 hours per day. 
The differential height measurements at the crossover points contain information on time-variable tidal displacement. 
Although most of the crossover points are concentrated in the polar regions due to the near-polar orbit, we can also expect many enough crossover points in the equatorial and low-latitude regions where the tidal amplitude is large, having the larger sensitivity for $h_{2}$. 
They assume in their simulation that $h_{2}$=1.3 (corresponding to the large $h_{2}$ case in Figure \ref{fig:h2}, and concluded that a total error of 2\% in $h_{2}$ is expected. On the other hand, for the large $h_{2}$ case, the $h_{2}$ accuracy required to distinguish presence or absence of a subsurface ocean is about 6\% (1$\sigma$) since $\sin$ (10 degrees) $\sim$ 0.17. 
Thus the GALA instrument specs and the observation plan will sufficiently confirm or disprove the existence of an ocean underneath ice shell.

\begin{figure}[!t]%
\centering
\includegraphics[width=70mm,clip]{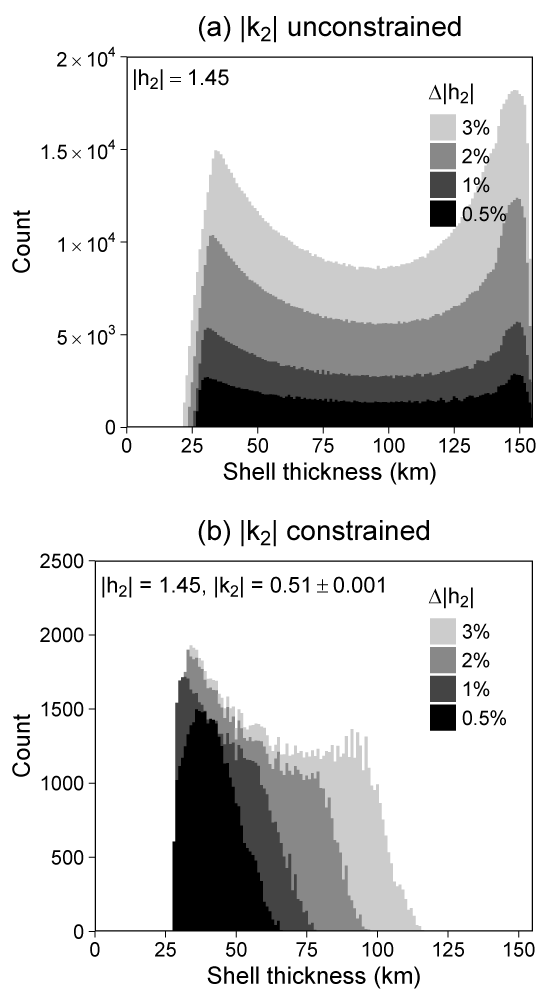}
\caption{Histogram of the thickness of the ice Ih shell for a given $|h_{2}|$ with different errors. Results under conditions where $|k_{2}|$ is (a) unconstrained and (b) constrained.}%
  \label{fig:shell_thickness}%
\end{figure}

Another implication which is scientifically important is the thickness of the top ice Ih shell (assuming a subsurface ocean exists) because this thickness yields not only the current interior thermal state but also the thermal evolution of the satellite. 
Figure \ref{fig:shell_thickness} indicates that a thinner ice shell leads to larger $|h_{2}|$ and $|k_{2}|$. 
Nevertheless, if one considers different physical properties of the shell (i.e., viscosity and rigidity), different thicknesses can lead to the same deformation amplitude (i.e., $|h_{2}|$ and $|k_{2}|$). 
Consequently, if one uses either of $|h_{2}|$ or $|k_{2}|$, the uncertainty in the shell thickness can be very large. 
For example, if the GALA data yield $|h_{2}|$ = 1.45, the thickness of the shell would be $\sim$30-150 km. 
While the phase lag also contains information of the shell thickness, a precise estimate of the shell thickness from the phase lag requires an extremely high precision for the phase lag
(i.e., \verb|<| 0.1 degree), which may not be practical for $|h_{2}|$.

However, the use of both Love numbers (i.e., $|h_{2}|$ and $|k_{2}|$) can reduce the uncertainty in the shell thickness significantly. For example, if $|k_{2}|$ is constrained to $\sim$0.51, the ice shell thickness is estimated to be $\sim$25-60 km. Thus, collaborative measurements and analyses of topography data and gravity field data are important for constraining the interior structure of Ganymede.

Tidal interaction also affects the moon's rotation, which depends on the interior structure especially on the existence of a subsurface ocean.
The rotational change of the moon statistically develops into a shifting of laser footprint position which could be detected with the GALA.
Furthermore combining altimetry data with visible imaging using camera (JANUS) leads to greater accuracy in detection of the rotational change.
Specifically, one of the rotational change, the forced libration, of Ganymede depends on the ice shell thickness, and its amplitude ranges between 15 meters for a 500 km-thickness ice shell and 355 meters for a 25 km-thickness ice shell \cite{baland10}. 
If Ganymede's ice shell is thin enough to enlarge the libration, it could be detected by the GALA altimetry.

\subsection{Small-scale roughness}
Slopes and roughness are important parameters for quantitative studies of surface characteristics. 
These parameters measure an interplay between various processes that modify a planetary surface, including crater formation, volcanic resurfacing and the generation of tectonics, and crustal strength and isostasy (see \cite{aharonson01} for Mars;\cite{cheng01,cheng02} for asteroid 433 Eros;\cite{barnouin08} for asteroid Itokawa;\cite{rosenburg11} for the Moon). 
To date the dominant factors affecting these parameters are known to vary with baseline scale and geologic environment especially. 
This is exemplified by surface slopes and roughness obtained at Mars by MOLA on board of MARS GLOBAL SURVEYOR.\cite{smith01} 
There remain unanswered questions (e.g.,\cite{rosenburg11}) on what causes some of the changes in broad scale surface roughness at broad baselines, which additional data collected on the icy surface of Ganymede could help elucidate. 
The GALA is perfectly suited to obtain these measurements.
The mean 100-m scale roughness of Mars
was determined to be 2.1 $\pm$ 2m RMS.\cite{garvin99} 
The GALA will have the capability to assess the surface roughness. 
The GALA will use both a digital filter but will also be capable of transmitting digitized return signals. With the latter, a better separation of albedo, surface roughness and local slope effects can be expected.
An important objective is to relate the surface roughness to different terrain types on Ganymede.\cite{schenk09} 
To characterize surface roughness at Ganymede laser footprints of 50 m and high shot frequencies will be used in the Ganymede orbit phases. 
Analysis of the digitized sample of the return pulse by the GALA provides the most-important data-sets for surface roughness. 

Local roughness, slopes and these distributions provide important clues to the morphologic history in terms of both formation and modification mechanisms.
Comparison of quantitative measurements of such roughness with surface geologic units is a powerful tool for interpreting the relationships between geologic and topographic units and their origins. 
Altimetric shots provide information characterizing the target surface within the laser footprint (diameter of 50 m at 500 km altitude orbit), since local roughness broadens the laser pulse and weakens the amplitude of the return signal.
Furthermore, statistical properties of local slopes and roughness provided by the GALA are also being used to characterize potential landing sites for future missions.

\subsection{Surface albedo}
Information on the surface albedo at the laser wavelength (1064 nm) can be gained from the intensities of the transmitted and returned pulses.
The GALA data also offer the advantage that no photometric (illumination) normalization is required to compare measurements of different areas of the moon's surface because the laser altimeter carries its own light source and will perform measurement at a constant phase angle.
Through mapping of these albedo patterns, possible correlations of albedo with topographic heights and geologic ages can provide information on geological processes and on the interaction of the surface with the moon's radiation and particle environment. 
Analysis of stereo topographic models obtained from Galileo images has revealed a strong correlation between elevations and material \cite{giese98, oberst99}. 
In some areas, bright material appears to be concentrated on slopes tilted north. 
Possibly, mass wasting in combination with illumination effects can explain these patterns. 
The GALA will map these albedo patterns and study their relationships to topographic elevations and slopes.
In addition, polar caps appear as bright regions in visible images at high latitudes which are connected directly with Jupiter's magnetic field lines and are therefore open to plasma bombardment from Jupiter's magnetosphere.
However the photometric behavior of Ganymede's surface has not yet been fully refined based on Galileo images and local reflectance is also affected by local morphology and geological processes. 
The GALA's albedo measurements will be a reliable tool in distinguishing between polar cap frost and other materials present elsewhere.
Such measurements would complement the data sets from the near-infrared spectrometers.
In addition, a laser altimeter can measure reflectance properties of unilluminated regions, e.g., permanent shadow area at polar locations, where camera data cannot be obtained.

\section{Possible Synergy of Other Scientific Instruments Onboard JUICE Spacecraft}
Characterization of the icy Galilean moons will be achieved not only from the GALA measurements but also by synergy of the various scientific instruments onboard JUICE spacecraft, for example, surface images taken by optical camera (JANUS) will confirm the position of the GALA laser footprint to complement the GALA point data for precise topographic mapping. 
A radar sounder (RIME) and a radio science experiment (3GM) probe the interior structure, especially the interior of the icy crust to investigate the formation of tectonic features.
A visible and infrared imaging spectrometer (MAJIS), an ultraviolet imaging spectrograph (UVS) and a sub-millimeter wave instrument (SWI) will acquire surface and atmosphere compositional data. 
A magnetometer (J-MAG) monitors moon's inductive response to the Jovian magnetic field and probes the subsurface ocean with the help of a particle environment package (PEP) and a radio and plasma wave investigation (RPWI). 
The instruments work closely together in a synergistic way with the GALA being one of the key instruments for understanding the evolution of the icy satellites Ganymede, Europa, and Callisto.

\section{Conclusion}

JUICE is a spacecraft mission to Jovian system, which is planned to be launched in 2022 aiming at coordinated observations of Jupiter and their icy Galilean moons especially Ganymede.
The GALA instrument is one of ten scientific payloads for JUICE and is currently under development with an international collaboration between European and Japanese institutions.
The GALA will be the first laser altimeter for icy bodies, and will measure the shape and topography of the Jovian icy Galilean moons, (globally for Ganymede, and flyby region for Europa and Callisto).
In previous observations, altimetry data were not obtained and image data has poor coverage for these moons because of only several flybys.
The GALA altimetry data will fill this gap and especially for Ganymede, the GALA altimetry at the 500 km altitude orbit around the moon allow an excellent global coverage of the entire Ganymede.
On a local scale, a first quantitative data provided by the GALA will significantly advance our understandings of the icy tectonics and of the small-scale surface conditions.
On a global scale, tidal responses, e.g., tidal deformation and rotational changes, of Ganymede can be detected and amplitudes bring us information on the presence or absence of the global liquid water ocean under the solid ice shell.
Furthermore, collaborative measurements with other scientific payloads will unveil the emergence of the potential deep habitat under the icy surface.

\section*{Acknowledgments}\label{Acknowledgments}
We thank an anonymous reviewer for constructive comments which greatly improved the manuscript.
This work was partially supported by the Japan Society for the Promotion of
Science (JSPS) KAKENHI grant 17K05635 (Jun Kimura) and 16K17787 (Shunichi Kamata).


\begin{thebibliography}{99}
\bibitem{grasset13}
Grasset, O., Dougherty, M.K., Coustenis, A., Bunce, E.J., Erd, C., Titov, D.,  et al.: JUpiter ICy moons Explorer (JUICE): An ESA Mission to Orbit Ganymede and to Characterise the Jupiter System, \textit{Planet. Space Sci.}, \textbf{78} (2013), 1--21.

\bibitem{hussmann13}
Hussmann H., Lingenauber, K., Michaelis, H., Kobayashi, M., Thomas, N., Lara, L., et al.: The Ganymede Laser Altimeter (GALA) as Part of the JUICE Payload: Instrument, Science Objectives and Expected Performance, European Planet. Sci. Congress, London, United Kingdom, 2013.

\bibitem{weiss04}
Weiss, J.W.: Planetary Parameters, \textit{Jupiter: The Planet, Satellites and Magnetosphere, Bagenal, F., Dowling, T. E., and McKinnon, W. B. (eds.)}, Cambridge University Press, London, 2004, pp.699--709.

\bibitem{anderson98}
Anderson, J. D., Schubert, G., Jacobson, R. A., Lau, E. L., Moore, W. B., and Sjogren, W. L.: Europa's Differentiated Internal Structure: Inferences from Four Galileo Encounters, \textit{Science} \textbf{281} (1998), pp.2019--2022.

\bibitem{pappalardo99}
Pappalardo, R., Belton, M. J. S., Breneman, H. H., Carr, M. H., Chapman, C. R., Collins, G. C., et al.: Does Europa Have a Subsurface Ocean? Evaluation of the Geological Evidence, \textit{J. Geophys Res.}, \textbf{104} (1999), pp.24015--24055.

\bibitem{zahnle03}
Zahnle, K., Schenk, P., Levison, H., and Dones, L.: Cratering Rates in the Outer Solar System, \textit{Icarus}, \textbf{163}, (2003), pp.263--289.

\bibitem{schmidt11}
Schmidt, B., Blankenship, D. D., Patterson, G. W., and Schenk, P. M.: Active Formation of 'Chaos Terrain' Over Shallow Subsurface Water on Europa, \textit{Nature}, \textbf{479}, (2011), pp.502--505.

\bibitem{kivelson00}
Kivelson, M. G., Khurana, K., K., Russell, C. T., Volwerk, M., Walker, R. J., and Zimmer, C.: Galileo Magnetometer Measurements: A Stronger Case for a Subsurface Ocean at Europa, \textit{Science}, \textbf{289}, (2000), pp.1340--1343.

\bibitem{anderson96}
Anderson, J.D., Lau, E.L., Sjogren, W.L., Schubert, G., and Moore, W.B.: Gravitational Constraints on the Internal Structure of Ganymede, \textit{Nature} \textbf{384} (1996), pp.541--543.

\bibitem{kivelson96}
Kivelson, M.G., Khurana, K.K., Russell, C.T., Walker, R.J., Warnacke, J., Coroniti, F.V., et al.: Discovery of Ganymede's Magnetic Field by the Galileo Spacecraft, \textit{Science} \textbf{384} (1996), pp.537--541.

\bibitem{kimura09}
Kimura, J., Nakagawa, T., and Kurita, K.: Size and Compositional Constraints of Ganymede's Metallic Core for Driving an Active Dynamo, \textit{Icarus}, \textbf{202}, (2009), pp.216--224.

\bibitem{sohl02}
Sohl, F., Spohn, T., Breuer, D., and Nagel, K.: Implications from Galileo Observations on the Interior Structure and Chemistry of the Galilean Satellites. \textit{Icarus} \textbf{157}, (2002) pp.104--119.

\bibitem{neukum97}
Neukum, G.: Bombardment history of the Jovian system, \textit{The Three Galileos: The Man, The Spacecraft, The Telescope, Barbabieri, C., Rahe, J., Johnson, T. V., and Sohus, A. M. (eds.)}, Springer, Dordrecht, 1997, pp.201--212.

\bibitem{neukum98}
Neukum, G., Wagner, R., Wolf, U., Ivanov, B.A., Head III, J. W., Pappalardo, R. T., et al.: Cratering Chronology in the Jovian System and Derivation of Absolute Ages, Lunar and Planetary Science Conference, Houston, U.S., 1998.

\bibitem{smith79a}
Smith, B. A., Soderblom, L.A., Johnson, T.V., Ingersoll, A.P., Collins, S.A., Shoemaker, E.M., et al.: The Jupiter System Through the Eyes of Voyager 1, \textit{Science}, \textbf{204}, (1979a) pp.951--957.

\bibitem{smith79b}
Smith, B. A., Soderblom, L.A., Beebe, R., Boyce, J., Briggs, G., Carr, M., et al.: The Galilean Satellites and Jupiter: Voyager 2 Imaging Science Results, \textit{Science}, \textbf{206}, (1979b), pp.927--950.

\bibitem{showman97} Showman, A. P. and Malhotra, R.: Tidal Evolution into the Laplace Resonance and the Resurfacing of Ganymede, \textit{Icarus}, \textbf{127} ,(1997), pp.93--111.

\bibitem{bland07} Bland, M. T. and Showman, A. P.: The Formation of Ganymede's Grooved Terrain: Numerical Modeling of Extensional Necking Instabilities, \textit{Icarus}, \textbf{189}, (2007), pp.439--456.

\bibitem{prockter01}
Prockter, L.: Icing Ganymede, \textit{Nature}, \textbf{410}, (2001), pp.25--27.

\bibitem{kivelson02}
Kivelson, M.G., Khurana, K.K., and Volwerk, M.: The Permanent and Inductive Magnetic Moments of Ganymede, \textit{Icarus}, \textbf{157} (2002), pp.507--522.

\bibitem{anderson01}
Anderson, J.D., Jacobson, R. A., McElrath, T. P., Moore, W. B., Schubert, G., and Thomas, P. C.: Shape, Mean Radius, Gravity Field, and Interior Structure of Callisto, \textit{Icarus}, \textbf{153} (2001), pp.157--161.

\bibitem{gao12}
Gao., P. and Stevenson, D. J.: Nonhydrostatic Effects and the Determination of Icy Satellites' Moment of Inertia, \textit{Icarus}, \textbf{226} (2012), pp.1185--1191.

\bibitem{map16}
Global Geologic Map of Ganymede, SIM3237, https://astrogeology.usgs.gov/ search/map/Ganymede/ Geology/Ganymede\_SIM3237\_Database (accessed Apr. 6, 2016).

\bibitem{palguta06}
Palguta, J., J.D. Anderson, G. Schubert, and W.B. Moore: Mass Anomalies on Ganymede, \textit{Icarus}, \textbf{180} (2006), pp.428-441.

\bibitem{palguta09}
Palguta, J., Schubert. G., Zhang. K., and Anderson, J. D.: Constraints on the Location, Magnitude, and Dimensions of Ganymede's Mass Anomalies, \textit{Icarus}, \textbf{201} (2009), pp.615-625.

\bibitem{giese08}
Giese, B., Wagner, R., Hussmann, H., Neukum, G., Perry, J., Helfenstein, P., et al.: Enceladus: An Estimate of Heat Flux and Lithospheric Thickness from Flexurally Supported Topography, \textit{Geophys. Res. Lett.}, \textbf{35(24)} (2008), L24204

\bibitem{golombek86}
Golombek, M. P. and W. B. Banerdt: Early Thermal Profiles and Lithospheric Strength of Ganymede from Extensional Tectonic Features, \textit{Icarus}, \textbf{68} (1986), pp.252-265.

\bibitem{nimmo02}
Nimmo, F., Pappalardo, R. T., and Giese, B.: Elastic Thickness and Heat Flux Estimates on Ganymede, \textit{Geophys. Res. Lett.}, \textbf{29(7)} (2002), pp.62-1--62-4.

\bibitem{pappalardo98}
Pappalardo, R.T., Head, J. W., Collins, G. C., Kirk, R. L., Neukum, G., Oberst, J., et al.: Grooved Terrain on Ganymede: First Results from Galileo High-Resolution Imaging, \textit{Icarus}, \textbf{135} (1998), pp.276-302.

\bibitem{driscoll94}
Driscoll, J. R. and Healy, D. M.: Computing Fourier Transforms and Convolutions on the 2-Sphere, \textit{Adv. Applied Mathematics}, \textbf{15(2)} (1994), pp.202--250.

\bibitem{kamata14}
Kamata, S. and Nimmo, F.: Impact Basin Relaxation as a Probe for the Thermal History of Pluto, \textit{J. Geophys. Res. Planets}, \textbf{119}, (2014),  pp.2272-2289. 

\bibitem{prockter10}
Prockter, L.M., Lopes,R.M.C., Giese, B., Jaumann, R., Lorenz, R. D., Pappalardo, R. T., et al.: Characteristics of Icy Surfaces, \textit{Space Sci. Rev.}, \textbf{153}, (2010), pp.63-111.

\bibitem{greeley00}
Greeley, R., Figueredo, P. H., Williams, D. A., Chuang, F. C., Klemaszewski, J. E., Kadel, S. D., et al.: Geological Mapping of Europa, \textit{J. Geophys. Res.}, \textbf{105} (2000), pp.22,559-22,578.

\bibitem{greenberg00}
Greenberg, R., Geissler, P., Tufts, B. R., and Hoppa, G. V.: Habitability of Europa's Crust: The Role of Tidal-Tectonic Processes, \textit{J. Geophys. Res.}, \textbf{105} (2000), pp.17551-17562.

\bibitem{steinbrugge15}
Steinbr\"ugge, G., Stark, A., Hussmann, H., Sohl, F., and Oberst, J.: Measuring Tidal Deformations by Laser Altimetry. A Performance Model for the Ganymede Laser Altimeter, \textit{Planet. Space Sci.}, \textbf{117} (2015), pp.184--191.

\bibitem{moore03}
Moore, W. B. and Schubert, G.: The Tidal Response of Ganymede and Callisto with and without Liquid Water Oceans, \textit{Icarus}, \textbf{166} (2003), pp.223--226.

\bibitem{wahr14}
Wahr, J.M. and Zhong, S.: The Effects of Laterally Varying Icy Shell Structure on the Tidal Response of Ganymede and Europa, \textit{J. Geophys. Res. Planets}, \textbf{119} (2014), pp.659--678.

\bibitem{kamata16}
Kamata, S., Kimura, J., Matsumoto, K., Nimmo, F., Kuramoto, K., and Namiki, N.: Tidal Deformation of Ganymede: Sensitivity of Love Numbers on the Interior Structure, \textit {J. Geophys. Res. Planets}, \textbf{121}, (2016), pp.1362--1375.

\bibitem{kamata15}
Kamata, S., Matsuyama, I. and Nimmo, F.: Tidal Resonance in Icy Satellites with Subsurface Oceans, \textit{J. Geophys. Res. Planets}, \textbf{120} (2015), pp.1528--1542.

\bibitem{baland10}
Baland, R--M. and Hoolst, T. V.: Librations of the Galilean Satellites: The Influence of Global Internal Liquid Layers, \textit{Icarus}, \textbf{209}, (2010), pp.651--664.

\bibitem{aharonson01}
Aharonson, O., Zuber. M. T., and Rothman, D.H.: Statistics of Mars' Topography from Mars Orbiter Laser Altimeter: Slopes, Correlations and Physical Models, \textit{J. Geophys. Res.}, \textbf{106}, (2001) pp.23723-23736.

\bibitem{cheng01}
Cheng, A.F., Barnouin-Jha, O., Zuber, M.T., Veverka, J., Smith, D.E., Neumann, G.A., et al.: Laser Altimetry of Small-Scale Features on 433 Eros from NEAR-Shoemaker, \textit{Science}, \textbf{292} (2001), pp.488--491.

\bibitem{cheng02}
Cheng, A.F., Barnouin-Jha, O.S., Prockter, L., Zuber, M.T., Neumann, G., Smith, D.E., et al.: Small-Scale Topography of 433 Eros from Laser Altimetry and Imaging, \textit{Icarus} \textbf{155} (2002), p.51--74.

\bibitem{barnouin08}
Barnouin-Jha, O.S., Cheng, A. F., Mukai, T., Abe, S., Hirata, N., Nakamura, R., et al.: Small-Scale Topography of 25143 Itokawa from the Hayabusa Laser Altimeter, \textit{Icarus}, \textbf{198} (2008), pp.108-124.

\bibitem{rosenburg11}
Rosenburg, M.A., Aharonson, O., Head, J. W., Kreslavsky, M. A., Mazarico, E., Neumann, G., et al.: Global Surface Slopes and Roughness of the Moon from the Lunar Orbiter Laser Altimeter, \textit{J. Geophys. Res.}, \textbf{116(E)} (2011), pp.E02001.

\bibitem{smith01}
Smith, D.E  Zuber, M. T., Frey, H. V., Garvin, J. B., Head, J. W., Muhleman, D. O., et al.: Mars Orbiter Laser Altimeter: Experiment Summary after the First Year of Global Mapping of Mars, \textit{J. Geophys. Res.}, \textbf{106} (2001), pp.23689-23722.

\bibitem{garvin99}
Garvin, J. B., Frawley, J. J., and Abshire, J. B.: Vertical Roughness of Mars from the Mars Orbiter Laser Altimeter, \textit{Geophys. Res. Lett.}, \textbf{26}, (1999), pp.381-384.

\bibitem{schenk09}
Schenk, P. M.: Slope Characteristics of Europa: Constraints for Landers and Radar Sounding, \textit{Geophys. Res. Lett.}, \textbf{36}, (2009), pp.L15204.

\bibitem{giese98}
Giese, B., Oberst, J., Roatsch, T., Neukum, G., Head, J. W., and Pappalardo, R. T.: The Local Topography of Uruk Sulcus and Galileo Regio Obtained from Stereo Images, \textit{Icarus}, \textbf{135} (1998), pp.303-316.

\bibitem{oberst99}
Oberst, J., Schreiner, B., Giese, B., Neukum, G., Head, J.W., Pappalardo, R.T., et al.: The Distribution of Bright and Dark Material on Ganymede in Relationship to Surface Elevation and Slopes, \textit{Icarus}, \textbf{140} (1999), pp.283-293.

\end{thebibliography}
\end{document}